\documentstyle[12pt,aaspp4,flushrt]{article}
\def\etal{{\it et~al.~}}
\def\Sec{${^{\prime\prime}}$\llap{.}}
\def\vmi{F606W--F814W~}
\renewcommand{\sun}{\odot}
\newcommand{\Msun}{\mbox{$M_\sun$}}
\input epsf

\begin{document}

\title{Hubble Space Telescope Observations of the Draco Dwarf
Spheroidal\footnote[1]{Based on observations with the NASA/ESA Hubble
Space Telescope, obtained at the Space Telescope Science Institute,
which is operated by AURA, Inc., under NASA contract NAS 5-26555.}}

\author{Carl J. Grillmair\footnote[2]{Department of Astronomy,
California Institute of Technology, Pasadena, CA 91125}, Jeremy R.
Mould\footnote[3]{Mount Stromlo and Siding Spring Observatories,
Institute of Advanced Studies, Australian National University, Weston
ACT 2611, Australia}, Jon A.  Holtzman\footnote[4]{Astronomy
Department, New Mexico State University, Box 30001, Dept. 4500, Las
Cruces, New Mexico 88003}, Guy Worthey\footnote[5]{Department of
Astronomy, University of Michigan, Ann Arbor, Michigan
48109-1090}\footnote[6]{Hubble Fellow},\\ Gilda E.
Ballester\footnote[6]{Department of Atmospheric, Oceanic, and Space
Sciences, University of Michigan, 2455 Hayward, Ann Arbor, MI 48109},
Christopher J. Burrows\footnote[7]{Astrophysics Division, Space
Sciences Department, ESA \& Space Telescope Science Institute, 3700
San Martin Drive, Baltimore, MD 21218}, John T.
Clarke\footnotemark[6], David Crisp\footnote[8]{Jet Propulsion
Laboratory, 4800 Oak Grove Drive, Pasadena, California 91109-8099},
Robin W. Evans\footnotemark[8], John S. Gallagher
III\footnote[9]{Department of Astronomy, University of Wisconsin --
Madison, 475 N. Charter St., Madison, WI 53706}, Richard E.
Griffiths\footnote[10]{Department of Astronomy, Johns Hopkins
University, 3400 N. Charles St., Baltimore, MD 21218}, J. Jeff
Hester\footnote[11]{Department of Physics and Astronomy, Arizona State
University, Tyler Mall, Tempe, AZ 85287}, John G.
Hoessel\footnotemark[9], Paul A. Scowen\footnotemark[11], Karl R.
Stapelfeldt\footnotemark[8], John T. Trauger\footnotemark[8], Alan M.
Watson\footnotemark[4], James A. Westphal\footnote[12]{Division of
Geological and Planetary Sciences, California Institute of Technology,
Pasadena, CA 91125}}

\begin{abstract}

We present an F606W-F814W color-magnitude diagram for the Draco dwarf
spheroidal galaxy based on {\it Hubble Space Telescope} WFPC2 images.
The luminosity function is well-sampled to $\sim 3$ magnitudes below
the turn-off. We see no evidence for multiple turnoffs and conclude
that, at least over the field of the view of the WFPC2, star formation
was primarily single-epoch. If the observed number of blue stragglers
is due to extended star formation, then roughly 6\% (upper limit) of
the stars could be half as old as the bulk of the galaxy. The color
difference between the red giant branch and the turnoff is consistent
with an old population and is very similar to that observed in the
old, metal-poor Galactic globular clusters M68 and M92. Despite its
red horizontal branch, Draco appears to be older than M68 and M92 by
$1.6 \pm 2.5$ Gyrs, lending support to the argument that the
``second parameter'' which governs horizontal branch morphology must
be something other than age.  Draco's observed luminosity function is
very similar to that of M68, and the derived initial mass function is
consistent with that of the solar neighborhood.

\end{abstract}


\section{Introduction.}

Galaxies like the Draco dwarf spheroidal may be the building blocks of
the Universe (Koo \etal 1994). Draco is among the faintest known
galaxies and indeed, at $2 \times 10^5 L\sun$, ranks among the
middling globular clusters in luminosity. However, its very extended
distribution of stars ($D \sim 0.5$ kpc) clearly distinguishes it from
the comparatively tightly condensed globulars. Moreover, assuming
dynamical equilibrium leads one to very high estimates for the
mass-to-light ratio (Aaronson 1983), with recent multifiber
spectroscopy yielding a remarkable $M/L = 90 \pm 15$ (Armandroff,
Olszewski, \& Pryor 1995).  If the dark matter is primarily stellar,
as might be inferred from the findings of recent microlensing
experiments, then the Initial Mass Function (IMF) in these galaxies
might have been very different from the present-day IMF in our own
Galaxy. It is clearly important to try to piece together the star
formation histories of such galaxies as this may have important
consequences for ideas concerning the formation and evolution of more
luminous galaxies like our own.

The color-magnitude (CM) diagram of the Draco dwarf was first studied
by Baade \& Swope (1961), and found to be generally similar to
the CM-diagrams of metal-poor globular clusters with the exception of
having a rather broad giant branch. Baade \& Swope also first detected
the ``anomalous'' Cepheids, more luminous than the period-luminosity
relation for type II Cepheids would predict, and now known to be
characteristic of dwarf galaxies. Another marked anomaly in the CM
diagram is Draco's red horizontal branch, which is incompatible with
low metallicity in a Population II system (the second parameter
problem).  This phenomenon alone could be taken to suggest that Draco
is $\sim2~\times~10^9$ years younger than classic, metal-poor globular
clusters like M92 (Lee, Demarque, \& Zinn 1994).  Other hypotheses
have been offered to explain the second parameter phenomenon (Renzini
1978), and an increasing body of evidence suggests that age may be no
more than a minor contributor to horizontal branch morphology
(Stetson, Vandenberg, \& Bolte 1996). However, the debate is by no
means settled, and the reader is referred to Lee \etal (1994) and
Chaboyer, Demarque, \& Sarajedini (1996) for arguments in support of the
age hypothesis.

The best ground-based CCD CM diagrams of the Draco dwarf
spheroidal to date have been presented by Stetson, Vandenberg, \&
McClure (1985) and Carney \& Seitzer (1986, hereafter CS86).  Like the
Ursa Minor dwarf spheroidal (the 2nd-faintest dwarf spheroidal known),
Draco appears to be as old as the oldest globular clusters, though
CS86 and others discuss evidence from blue stragglers for an
intermediate-age population as well.
 
  The superb resolution afforded by the Wide Field/Planetary Camera 2
(WFPC2) on the {\it Hubble Space Telescope} permits a remarkable
improvement in our ability to study the stellar content of the Draco
dwarf spheroidal as well as other galaxies in the Local Group. Here we
describe the first results of an {\it HST} program to investigate the
stellar population of this galaxy.

\section{Observations and Photometry. \label{sec:obs} }

The images we presently discuss were taken on June 9, 1995 with WFPC2.
The region selected for WFPC2 imaging is shown overlaid on the Digital
Sky Survey in Figure \ref{fig:plate1} (Plate 1), and is roughly
centered on Field 1 of CS86.  One 200s and two 1000s exposures were
taken in F606W ($\sim V$), while the F814W ($\sim I$) images comprise
200s, 1100s, and 1300s integration time. F606W was chosen over F555W
to give a somewhat higher signal-to-noise ratio. In the process of
cosmic ray removal, the exposures in each filter were combined into
single, composite images from which subsequent photometry was
derived. A mosaic of the coadded WFPC2 frames in F814W is shown in
Figure \ref{fig:plate2} (Plate 2).

Photometry on the three Wide-Field (WF) detectors (0\Sec0996
pix$^{-1}$) was carried out using a combination of aperture photometry
and ALLFRAME PSF-fitting photometry (Stetson 1994).  ALLFRAME is the
latest development in the DAOPHOT series of photometry packages and
differs from its predecessors primarily in its ability to use
information from many individual frames simultaneously to better
constrain the final fitted magnitudes. ALLFRAME is applied to the data
{\it after} object detection and aperture photometry is carried out
using DAOPHOT II (Stetson 1987).

To push the star detection limit to as faint a level as possible, we
coadded all images taken through the {\it same} filter.  Coadding the
F606W and F814W images prior to object detection might in principle
seem to offer an even fainter detection limit, but doing so would have
severely compromised the reliability of subsequent completeness
determinations. Two DAOPHOT detection passes were carried out,
separated by photometry and subtraction of all stars found in the
first pass. Faint stars hidden within the PSF skirts of brighter
companions were thereby revealed and subsequently added to the list of
detected stars to be measured.  Point-spread functions for each
detector and each filter were constructed within DAOPHOT using the
$\sim 20$ brightest and most isolated point sources in each frame.
ALLFRAME was then applied to the F606W and F814W images
simultaneously.  A total of 4429 stars were ultimately detected and
measured in the three WF frames.

Due to the degree of undersampling and relatively poor PSF
representation in the WF chips, PSF-fitting of brighter stars is
subject to greater uncertainties in the absence of crowding than is
pure synthetic-aperture photometry. However, crowding and poor
signal-to-noise ratio become ever increasing problems as one considers
fainter and fainter stars, and the ALLFRAME-derived magnitudes yield a
tighter main sequence than is obtainable using aperture photometry.
In order to retain the small uncertainties associated with aperture
photometry of bright, uncrowded giants and to simultaneously extend
the main sequence as far as possible, the color-magnitude diagram was
constructed using both aperture and PSF-fitted magnitudes. The
aperture magnitudes were used whenever the predicted total light
within an aperture of radius 2 pixels from all other resolved stars
(based on the known characteristics of the WFPC2 PSF) within
2$^{\prime\prime}$ was less than 1\% of the light from the star
centered in the aperture. Corrections to a 0\Sec5 aperture were made
in each case, and the synthetic zero points of Holtzman \etal (1995)
were used to compute WFPC2-system magnitudes.

Completeness tests were carried out by adding $\approx 100$ stars at a
time to each of the F606W and F814W images. Stars were added in
integer--F814W steps, and with F606W - F814W $\approx 0.9$ mag.  The
frames were then processed using DAOPHOT II and ALLFRAME in a manner
identical to that applied to the original data.  The results of these
tests are shown in Figure \ref{fig:completeness}, which shows that the
50\% completeness level occurs at F814W $\approx 26.2$. The faintest
stars we can reliably measure therefore have masses of $\sim 0.6
M_{\odot}$. The tests also show that the photometric uncertainties are
$\pm0.06$ mag rms at F814W = 23, and $\pm0.11$ mag rms at F814W = 25.

\section{Discussion \label{sec:discussion}}

\subsection{Morphology of the Color-Magnitude Diagram}

The F814W vs F606W--F814W color-magnitude (CM) diagram, constructed
using both aperture photometry for uncrowded stars (269) and ALLFRAME
fits for crowded stars (4156) is shown in Figure \ref{fig:cm}.  An
electronic version of the photometry table is available on request
from CJG.

The morphology in Figure \ref{fig:cm} above the turnoff is very
similar to that found by Stetson \etal (1985) and CS86.  Owing to the
superior resolution afforded by the {\it HST} data, we are able to
extend the main sequence 2 magnitudes fainter than these ground-based
studies. The turnoff region is well resolved and reveals that the
bluest main-sequence stars have \vmi = 0.36. There are evidently fewer
blue stragglers than in the CM diagram of CS86, consistent with a
ratio of $\sim 5$ in field of view. Unlike CS86, we see no evidence
for distinct multiple turnoffs; most of the stars within the WFPC2
field of view appear to be approximately coeval. We caution however
that, while our photometry may be more accurate, we have sampled only
one fifth as many stars in the turnoff region as have CS86, and it may
be premature to rule out the multiple turnoffs in Draco. Additional
WFPC2 fields of Draco will be required before we can seriously
address this issue.

\subsection{Age}

In Figure \ref{fig:cmtracks} we compare the Draco CM diagram with
fiducial sequences for the Galactic globular clusters M68 (Walker
1994) and M92 (Heasley \& Janes 1996), which have [Fe/H] = -2.09 and
-2.24 dex, respectively (Djorgovski 1993).  The Draco magnitudes have
been dereddened assuming E({\it B-V}) = 0.03 (Stetson 1979a) using the
absorptions tabulated by Holtzman \etal (1995) for stars of K5
spectral type. The globular sequences were dereddened and translated
using the color excesses and distances tabulated by Peterson (1993). A
mean abundance for giants in Draco has been spectroscopically measured
to be [Fe/H] $\approx -1.9$ dex (Lehnert \etal 1992).  Adopting a mean
horizontal branch magnitude $\langle m_V \rangle = 20.07 \pm 0.03$
(Stetson 1979a), E$(B-V)$ = 0.03 (Stetson 1979b), and $M_V$ = 0.15
[Fe/H] + 0.82 (Carney \etal 1992), we use $(m-M)_0 = 19.48.$ The
isochrones shown in Figure \ref{fig:cmtracks} are from Vandenberg \&
Bell (1985), with colors from the $V-K$ vs $T_{eff}$ relations for
dwarfs of Johnson (1966), and for giants of Ridgway \etal
(1980). $V-K$ colors have been transformed to $V-I$ empirically and in
the correct metallicity regime via the preliminary multimetallicity
color-color diagrams of Worthey \& Fisher (1996), and $V-I$ have in
turn been converted to F606W--F814W using the transformation equations
in Holtzman \etal (1995).

A sequence of normal points representing the color-magnitude sequence
was generated by defining an inclusion envelope around the main
sequence, sub-giant, and giant branches, and computing mean magnitudes
and colors in 0.4 magnitude-wide bands. Due to the scarcity of giant branch 
stars and the associated large uncertainties in their mean
colors, the locus of the giant branch above F814W = 20.5 was estimated
by eye. This sequence, resampled using spline fits, was then compared
using a maximum-likelihood technique with isochrones ranging in age
from 12 to 18 Gyrs, and in [Fe/H] ranging from -1.2 to -2.2. The
normal points were uniformly weighted, and the best-fitting isochrone
had an age of 16 Gyrs and [Fe/H] = -2.2. Given the weighting of the
data, a formal $\chi^2$ estimate of the uncertainty was not
straightforward, and the fitting uncertainty was instead estimated by
generating one hundred realizations of the sequence of normal points
(using Gaussian deviates based on the dispersion in the mean colors in
each magnitude bin) and comparing them to each of the isochrones. This
yielded an age uncertainty, based purely on random and fitting errors,
of 0.5 Gyr.

Shown in Figure \ref{fig:cmtracks} are isochrones corresponding to an
age of 16 Gyr and [Fe/H] ranging from -1.2 to -2.2 dex. An age of 16
Gyrs is not atypical of ages found for several metal-poor Galactic
globular clusters using this set of isochrones. However, despite the
small uncertainty due to random effects, it is apparent from Figure
\ref{fig:cmtracks} that there are clear, systematic differences
between the isochrones and the data. Whereas the fiducial sequences of
M68 and M92 agree reasonably well with the CM distribution of Draco
stars, the isochrones generally appear too red, particularly along the
giant branch. This is undoubtedly due in large part to the roundabout
technique used above to transform values of $T_{eff}$ to F606W--F814W
colors.  Given the numerous calibrations involved and the systematic
discrepancies which can arise, it is clearly dangerous to place much
faith in the age we derive using this method.

A {\it calibration-independent} method for measuring relative
ages\footnote{We emphasize that our conclusions refer to {\it
relative} ages. If RR Lyrae stars are as luminous as claimed by Reid
(1997) and Alcock \etal (1997) (M$_V = +0.20$), then the {\it
absolute} ages of the globular clusters (and Draco) are reduced by
$\sim 3$ Gyrs.} has been described by Vandenberg, Bolte, \& Stetson
(1990). This involves measuring the difference in color between the
turnoff and the giant branch, which is relatively insensitive to
metallicity but is a monotonic function of age.  In Figure
\ref{fig:cmaligned} we show the fiducial sequence for Draco (generated
as described above) compared with similar sequences for the Galactic
globular clusters M68 and M92.  Also shown are isochrones having
[Fe/H] = -2.0 and ages ranging from 13 to 18 Gyr. The globular cluster
sequences and the isochrones have been shifted horizontally to match
the Draco sequence at the bluest point of the turnoff region, and
vertically to agree with one another at a point 0.05 magnitudes
redward of the turnoff on the main sequence. The Draco sequence
evidently agrees very well with the globular cluster sequences,
though there are small differences in the shape of the turnoff
and the slope of the giant branch. 

   The relative ages between sequences are best measured in this
technique by comparing the colors of the giant branch, where the color
separation is largly independent of magnitude. Looking at the
isochrone spacing, it is evident that the color difference between
giant branches is not uniform, but depends on absolute age. If we
calibrate age differences using the oldest (leftmost) two isochrones,
then the mean color difference measured in the region 20 $<$ F814W $<$
21 (below the HB and in a region where the giant branch is relatively
well sampled) yields a relative age difference between Draco and
either of the globular clusters of considerably less than one
Gyr. Note that using increasingly younger isochrones to establish the
color difference--age calibration would reduce the inferred age
differences even more. 

Given the photometric scatter apparent on the main sequence, the
presence of blue stragglers or large numbers of binary stars could
conceivably affect the determination of the mean colors of the turnoff
and the main sequence. To estimate our sensitivity to these effects,
we have recomputed the Draco normal points using an inclusion envelope
alternately more exclusive of stars on the blue side of the turnoff
and on the red side of the main sequence. In the latter case, we have
excluded approximately three times as many outliers to the red of the
main sequence as to the blue.  Using these computed normal sequences
(in addition to a number of hand-sketched realizations) and bringing
all other sequences into alignment as before, we estimate our
measuring uncertainty to be $\pm 1$ Gyr. The distribution of relative ages
measured in this way is slightly asymmetric in the sense that an older
Draco is preferred over a younger.

Additional uncertainty arises from possible errors in the color terms
used to transform between photometric systems.  Two such color terms
are used here: the term used to transform the ground-based data into
V-I, and that used to transform V-I to F606W-F814W. The close
similarity in age we infer for M68 and M92 agrees with the results of
Vandenberg et al. (1990) based on independent B, V data sets and
suggests that errors in the ground-based color terms are likely to be
small.  Errors in the transformation from V-I to F606W-F814W arise
from inaccuracies in the fitting formula used on the synthetic
photometry results and from the error made by applying a
transformation based on stars of solar metallicity to a metal-poor
population. We estimate that fitting errors would introduce an error
of order 0.02 magnitudes in the difference in color between the
turnoff and the giant branch.  This translates to an age uncertainty
of about 2.3 Gyrs. 

Applying transformation coefficients based on solar-metallicity stars
to the more metal-poor stars of M68 and M92 has the effect of making
the globular cluster giant branches in Figure \ref{fig:cmaligned}
appear somewhat bluer relative to Draco than they should. Using Kurucz
model atmospheres, we find that stars at the turnoff and on the giant
branch are fainter in F606W by 0.014 and 0.024 magnitudes,
respectively, than what we compute using the Holtzman \etal (1995)
coefficients. The F814W transformation for these stars is much less
sensitive to metallicity, and we estimate that the differences in
F606W-F814W color between the turnoffs and the giant branches of M68
and M92 should be about 0.014 magnitudes larger than shown. Correcting
our measured result accordingly and combining the two major
sources of error above, we conclude that Draco is $1.6 \pm 2.5$ Gyrs
older than either M68 or M92.

Given the very different horizontal branch morphologies of Draco and
the two globular clusters, what can we say about the 2nd parameter
problem? In Figure \ref{fig:ldz} we partially reproduce the [Fe/H]
vs. HB morphology diagram of Lee \etal (1994). We have determined the
morphological type quantity (B-R)/(B+V+R) for Draco using the
color-magnitude diagrams of both Stetson (1979a) and CS86. Counting
stars on the blue and red sides of the instability strip as well as
likely RR Lyrae variables, we find (B-R)/(B+V+R) = -.77 and -0.70 for
the two data sets, respectively. We plot the mean of these two values
in Figure \ref{fig:ldz} along with similar quantities tabulated by Lee
\etal (1994) for M68 and M92. For both uniform mass-loss models and
models in which the mass-loss rate increases with age (Reimers 1977),
the Draco point appears to fall about 2 Gyr below the isochrone which
best matches M68 and M92. As this is only 1.4$\times$ our estimated
age uncertainty, we cannot completely rule out consistency between
Draco and an age-dependent HB morphology of the type modeled by Lee
\etal.  WFPC2 photometry of old globular clusters, sufficient to
reduce the relative age uncertainty to ~$\sim 0.5$ Gyr, will be
required before we can put firmer constraints on the models.

\section{Star Formation History}

Zinn (1978) and Lehnert \etal (1992) have demonstrated an abundance
spread of approximately 1 dex in Draco. Aside from this chemical
inhomogeneity, how well does the Draco dwarf spheroidal galaxy fit the
{\it simple stellar population} paradigm of Renzini \& Buzzoni (1986)?
We have seen that the turnoff CM diagram morphology of Draco is very
similar to that of M92; globular clusters are the prototype simple
stellar populations, coeval systems of initially homogeneous
composition.  Clearly, Draco is predominantly a simple stellar
population originating as early as anything in the Galaxy.

Close examination of the CM diagram shows a significant population of
blue stragglers (see also CS86).  Such stars are often seen in simple
stellar populations, including both ancient globular and younger open
star clusters, but they can also result from an $extended$ star
formation history.  We can count them in color profiles (Figure
\ref{fig:color}) formed out of half-magnitude strips from the turnoff
region.  The surface density of foreground stars in this region of the
CM diagram (specifically a region 0.3 magnitudes wide in color and 2
magnitudes high in F606W) is predicted by the Galactic model of
Bahcall and Soneira (1980) to be 17 per square degree. Thus, there is
only a 2\% probability of finding a single foreground star among the
$\sim 17$ stars we identify as blue stragglers (roughly speaking,
those stars with F814W $< 23$ and \vmi $< 0.3$). There are between 60
and 70 stars in the same region ($V < 23$ and $B-V < 0.3$) of the CM
diagram of CS86.  Given that the area surveyed by CS86 is about 5
times that of the area covered by our WFPC2 field, a ratio of $3.8 \pm
1$ in the number of blue stragglers indicates reasonable agreement.

Simulations of the turnoff region using the Yale isochrones (Green,
Demarque \& King 1984) require 6\% of the mass to be contained in a
second population, half the age of the primary population, in order to
fully account for the blue stragglers in Figure \ref{fig:cm}. We note
that 6\% is an upper limit on the size of a second younger component
of Draco, since blue stragglers can be an intrinsic part of a simple
stellar population.

A putative extended star formation history strengthens the link
between Draco and more fully developed examples of this phenomenon,
such as Carina (Smecker-Hane \etal 1994; 1996). The difference between
Draco and Carina and galaxies such as LGS 3 (Mould 1997) which are
still forming stars today (at 10\% of the original rate in the latter
case) is then simply one of degree. Draco simply had a shorter tail to
its initial burst of star formation. There is considerable evidence
that wind-induced mass-loss cannot account for the stellar populations
seen in several Local Group dwarfs, and that stripping or capture of
gas must have occurred (see Gallagher \& Wyse 1994 for a review).
Draco is currently among the nearest of the dwarf galaxies and, if its
orbit is largely contained within its present distance, the
probability of capturing significant amounts of rejuvenating
circumgalactic gas may be substantially smaller than for dwarfs in the
outskirts of the Local Group. Alternatively, a high ultraviolet flux
from the precursors to the Galaxy and M31 may have removed substantially more 
gas from the nearby dwarfs (van den Bergh 1994).

None of these remarks bear at all on the better-known curious
phenomenon of Draco's red horizontal branch. The finding that Draco is
as old or older than M92 requires us to focus on the problem of
understanding Draco's RHB. Higher S/N (and preferably F555W--F814W)
photometry performed differentially with respect to [Fe/H] = --2
clusters have the potential to sharpen our conclusions further. If the
relative age uncertainty could be reduced to 0.5 Gyr, one would have
to turn to parameters such as cluster density (e.g. Fusi Pecci \etal
1996) as the second parameter in the globular cluster system. The age
spread of the halo, which is the defining feature of the Searle \&
Zinn (1978) galaxy formation scenario (as opposed to Eggen,
Lynden-Bell, \& Sandage 1962) would then need to be defined by the
ages of the low density element of the halo, namely Draco, its sibling
dwarf spheroidals, and the Galactic bulge.

\subsection{The Initial Mass Function}

Figure \ref{fig:lf} shows the differential luminosity function (LF)
for the stars in our sample. Also shown is a LF computed from Walker's
(1994) photometry of M68. No completeness corrections have been
applied to the M68 data, and the counts have been arbitrarily
normalized to our data at F814W = 23. Note that the number of measured
stars in the M68 sample is about $6\times$ larger than in the Draco
sample.  The Draco LF appears to be somewhat steeper on the giant
branch, and somewhat shallower below the turnoff.  Uncertainties
introduced by differences in noise characteristics, resolution, and
detection efficiency make it difficult to draw firm conclusions but,
over the (admittedly narrow) mass range considered here, there does
not appear to be an obvious bias towards lower-mass stars which might
be held to account for the very high estimates of Draco's
mass-to-light ratio.  Based on the present data set, all we can say is
that if the bulk of the matter is in the form of stars, such stars
must have masses less than $0.6 M_{\odot}$.

The observed LF can be used to place limits on the
initial mass function (IMF) in Draco. The current observations are of
stars with masses between 0.6 and 0.9 \Msun.  We parameterize the IMF
in this mass range by a simple power law, with $dN/dM\propto
M^\alpha$.  Even stars of this low mass evolve significantly over a
Hubble time, so the inferred IMF slope depends on the age of the
stellar population in Draco.

We construct model LFs using a variety of IMFs and
compare each model distribution with the observed one to see if it is
plausible that they are drawn from the same parent population. The
construction of model LFs allows us to incorporate photometric errors
(both random and systematic) and incompleteness into the analysis; the
incompleteness and errors which we apply to the model LFs are those
determined from our artificial star experiments.  We can also include
the effect of a binary star contribution if we make an assumption about
the frequency and distribution of the relative masses of stars in
binary pairs. For Draco, we assume a metallicity of [Fe/H]=-2, a distance
and extinction as quoted above, and use Vandenberg (1983) isochrones 
(from which Worthey's models are derived for low mass stars) along with
Kurucz model atmospheres to convert from masses to assumed luminosities.

We define as unacceptable IMF slopes for which the model LF and
the observed LF are inconsistent with each other at the 95\% confidence
level as determined by a Kolmogorov-Smirnoff test.  With this
definition, we constrain the IMF slope to fall within the range
\begin{description}
\item{$\phantom{-0.00>}\alpha>-1.00$ for an assumed age of 18 Gyr,}
\item{$-1.52>\alpha>-1.98$ for an assumed age of 15 Gyr,}
\item{$-2.14>\alpha>-2.30$ for an age of 12 Gyr,  and}
\item{$-2.48>\alpha>-2.56$ for an age of 10 Gyr}
\end{description}
when we fit the F606W LF over the range
$1<M_{F606W}<7.25$ with stellar models of a single metallicity,
[Fe/H]=-2.  The dependence on age is easy to understand: as a
population evolves, the LF around the turnoff
steepens, leading to a flatter inferred IMF for an older population.
These results were derived assuming a binary fraction of 0.5 with
uncorrelated masses of the binary stars; with no binaries, we get
slopes which are slightly shallower.  Similarly, small variations on
the above allowed ranges are derived with different assumptions about
the metallicity, distance, and/or a different adopted magnitude range
over which the comparison with the observed LF is made.

If we fit the F814W LF, we infer a range of possible
IMF slopes which are slightly steeper than those inferred from the
F606W LF. For no binaries, the inferred slope range
overlaps between the two bandpasses; for a binary fraction of 0.5, all
of the acceptable IMF slopes for F814W are steeper than those allowed
by the F606W LF.

For comparison, the IMF in this mass range in the solar neighborhood
is proportional to $M^{-2.2}$ (Kroupa, Tout, \& Gilmore 1993). The
allowed slopes for Draco are remarkably similar to this value.

\section{Conclusions}

Based on photometry of WFPC2 images of the Draco dwarf spheroidal, we
conclude the following:

\begin{itemize}

\item Draco is predominantly an old system, being $1.6
\pm 2.5$ Gyrs older than either M68 or M92.

\item We see no evidence for multiple turnoffs which would be
indicative of episodic star formation.

\item The presence of significant numbers of blue stragglers {\it may}
indicate an extended period of star formation, though the numbers of
younger stars $< 6$\% of the first generation.

\item The luminosity function looks in all respects nearly identical to that
of the old globular cluster M68.

\item The slope of the initial mass function appears to be very
similar to that of the IMF determined for the solar neighborhood.

\end{itemize}

Future work should be aimed at refining the relative ages of Draco and
blue horizontal branch clusters like M92. Obtaining both higher
signal-to-noise ratio data as well as WFPC color-magnitude diagrams of
metal poor globular clusters should allow us to reduce the relative age
uncertainty to $\approx 0.5$ Gyr. Such data would considerably
strengthen constraints on the process and mechanisms which led to the
formation of our Galaxy.

\acknowledgments

We are grateful to Gary da Costa for a critical reading of an earlier
version of this manuscript. This research was conducted by the WFPC2
Investigation Definition Team, supported in part by NASA Grant
No. NAS7-1260.

\clearpage

\figcaption[grillmair.fig1.ps]{Digitized Sky Survey image of the field
containing the Draco dwarf spheroidal.  The outlines indicate the HST
WFPC2 field and Fields 1 and 2 of Carney and Seitzer (1986). The
entire field shown subtends $15^{\prime}$ on a
side.\label{fig:plate1}}

\figcaption[grillmair.fig2.ps]{Mosaic of the coadded F814W
frames. \label{fig:plate2}}

\figcaption[grillmair.fig3.ps]{Completeness fraction as a function of
F606W and F814W magnitude. Completeness tests were made using
artificial stars of color F606W - F814W $\approx 0.9$
mag.\label{fig:completeness}}

\figcaption[grillmair.fig4.ps]{HST color-magnitude diagram of the
Draco dwarf spheroidal.  Most of the stars on the giant branch were
measured using aperture photometry, while the majority of stars below
the turnoff were measured by fitting PSFs.  \label{fig:cm}}

\figcaption[grillmair.fig5.ps]{The color-magnitude diagram of the
Draco dwarf spheroidal compared with fiducial sequences for the
metal-poor globular clusters M68 and M92. Superimposed are Worthey
models of age 16 Gyrs and spanning metallicities of [Fe/H] = -2.2 on
the left to [Fe/H] = -1.2 on the right. The globular cluster sequences
have been transformed from $V$ and $I$ to F606W and F814W using the
transformation coefficients given by Holtzman \etal
1995.\label{fig:cmtracks}}

\figcaption[grillmair.fig6.ps]{The fiducial sequence for Draco
compared with similar sequences for the Galactic globular clusters M68
and M92, using the relative-age-measuring technique described by
Vandenberg, Bolte, \& Stetson (1990). The M68 and M92 sequences,
measured in $V$ and $I$, have been transformed to F606W, F814W using
the synthetic transformation discussed in the text. The cluster
sequences and the Worthey models (having [Fe/H] = -2.0 and ages
ranging from 13 Gyr on the right to 18 Gyr on the left) have been
shifted horizontally to match the Draco sequence at the bluest point
of the turnoff region, and vertically to agree with one another at a
point 0.05 magnitudes redward of the turnoff on the main sequence. The
filled circles correspond to the locus of the giant branch determined
as described in the text. \label{fig:cmaligned}}

\figcaption[grillmair.fig7.ps]{A partial reproduction of Lee \etal's
(1994) Figure 9a, which they use to demonstrate a relation between
horizontal branch morphology and age. The solid lines correspond to
models which assume uniform stellar mass loss and are separated by 2
Gyrs. The dashed lines are also separated by 2 Gyrs and assume a
mass-loss rate which increases with age. The plotted values for
various globular clusters are taken from the compilation of Lee \etal
(1994), and the HB type for Draco was computed using the
color-magnitude diagrams of Stetson (1979a) and Carney \& Seitzer
(1986).\label{fig:ldz}}

\figcaption[grillmair.fig8.ps]{Color profiles generated by counting
stars lying within half-magnitude-wide F814W bins. We attribute the
excess number of blue stars over a putative Gaussian color
distribution to the presence of $\sim 17$ blue
stragglers. \label{fig:color} }

\figcaption[grillmair.fig9.ps]{The differential luminosity function of
the stars in Draco. The error bars reflect Poisson errors only, and
the dashed histogram shows the effect of dividing by the completeness
function.  Note that no account has been taken of bin-jumping due to
photometric errors. The dotted line shows a luminosity function
derived from Walker's M68 photometry {\it without} having applied
completeness corrections. \label{fig:lf} }

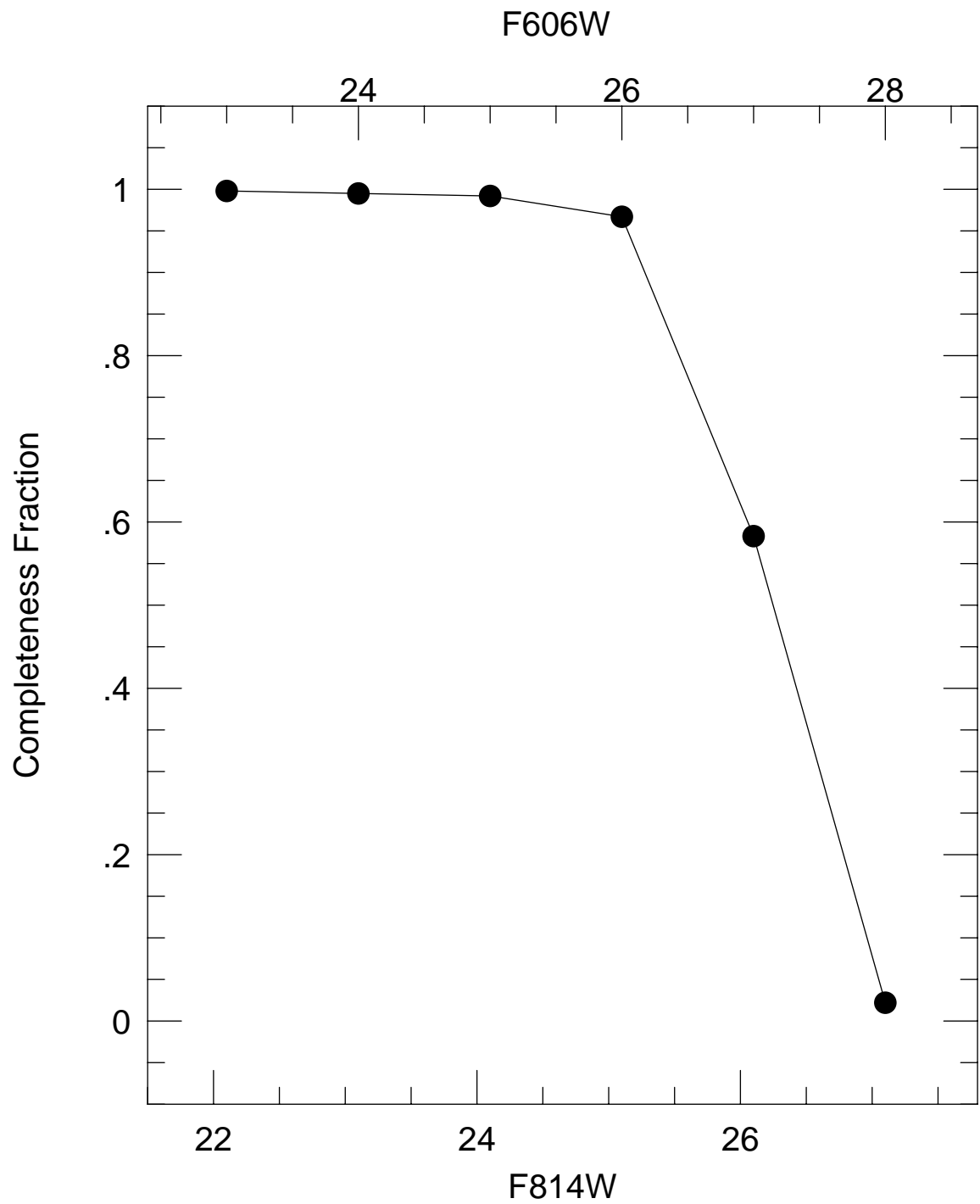
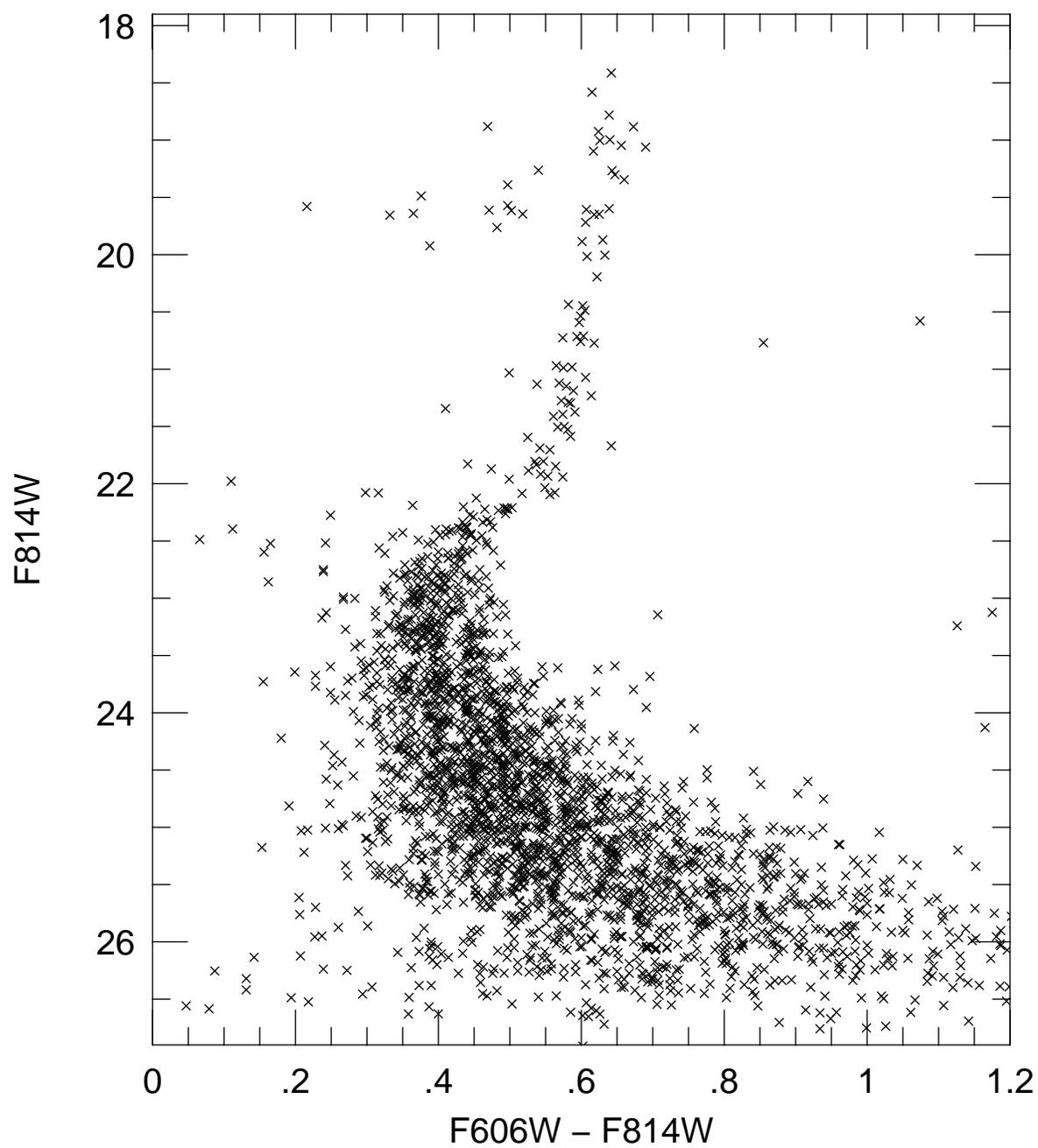
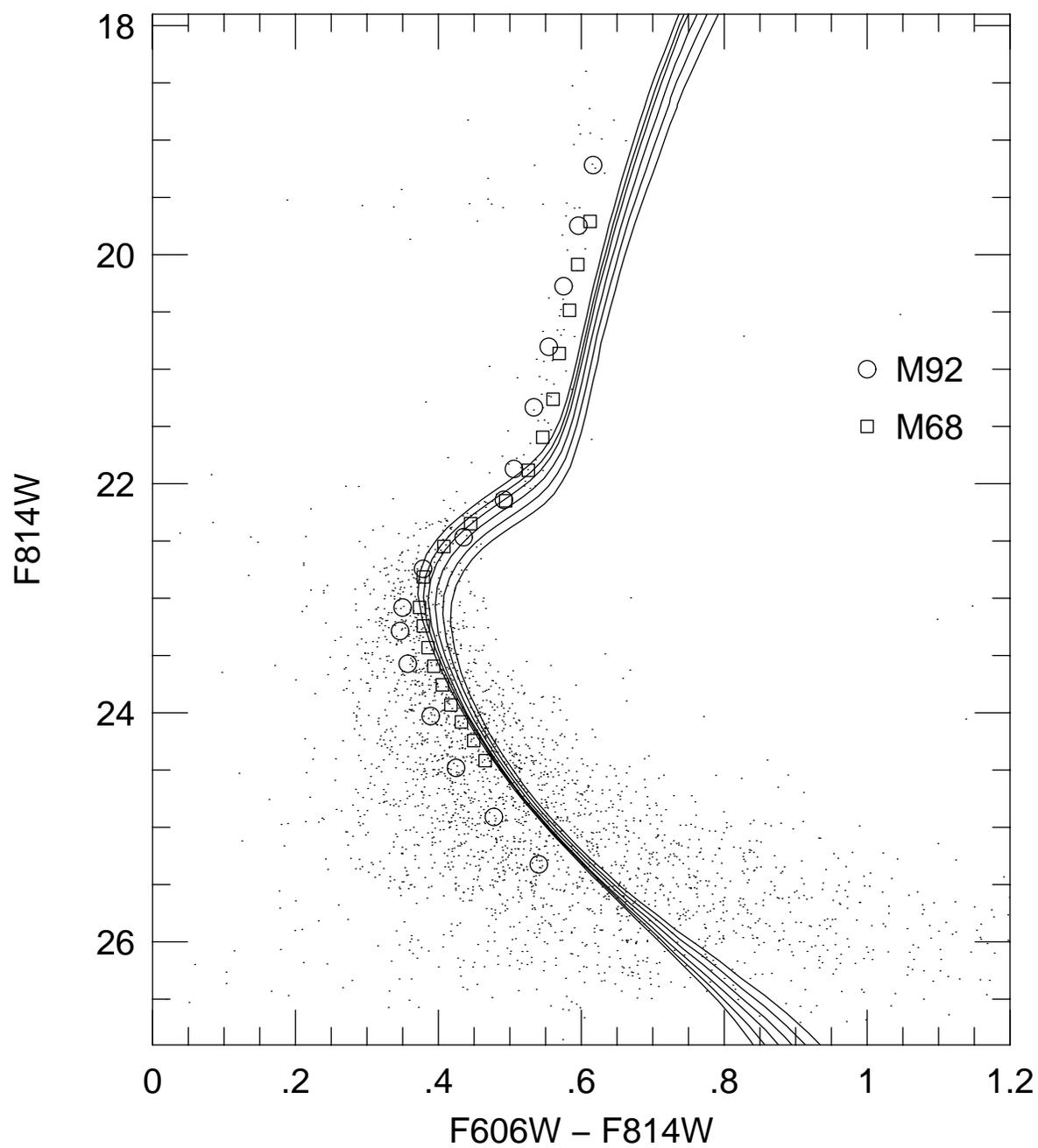
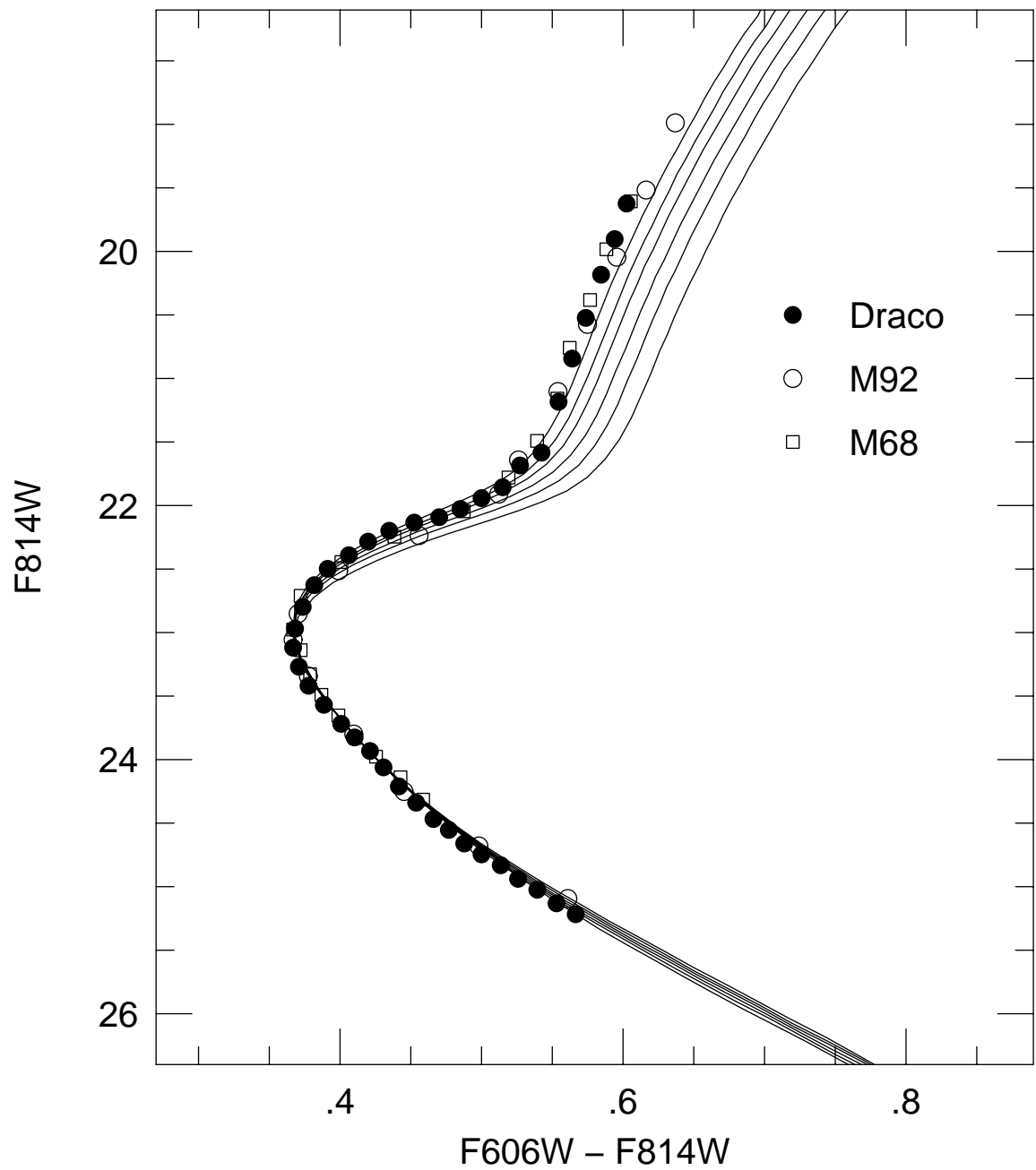
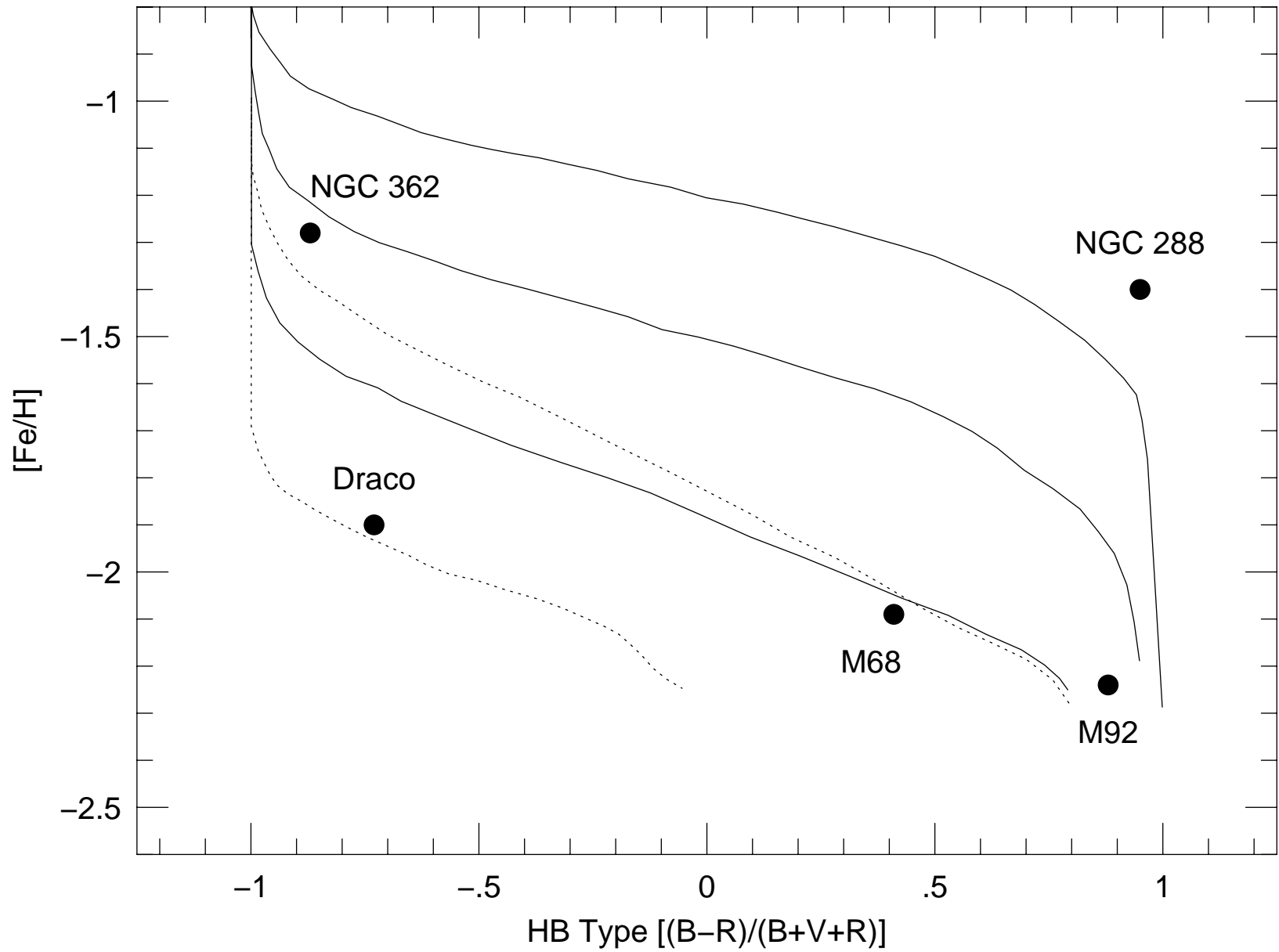
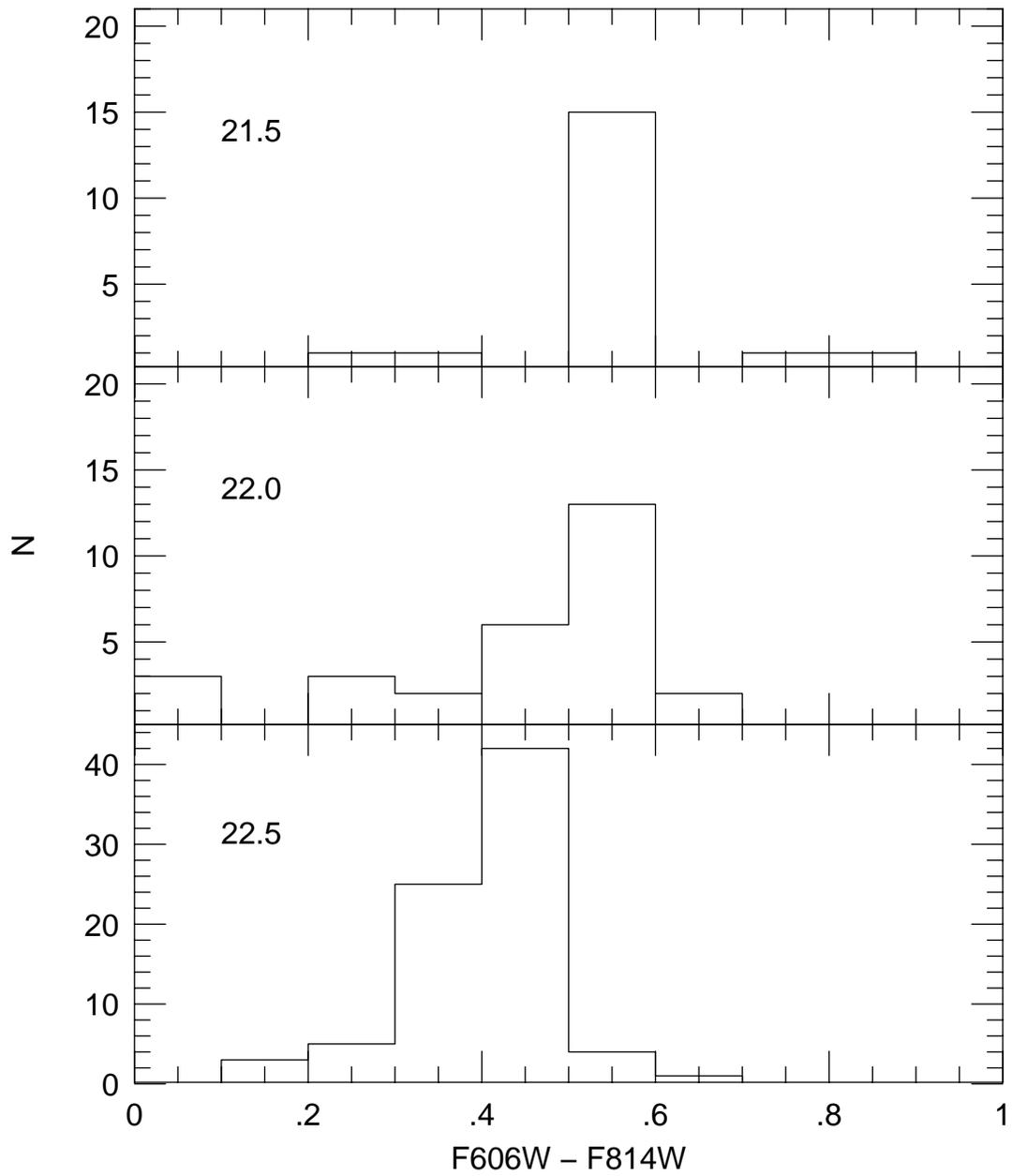
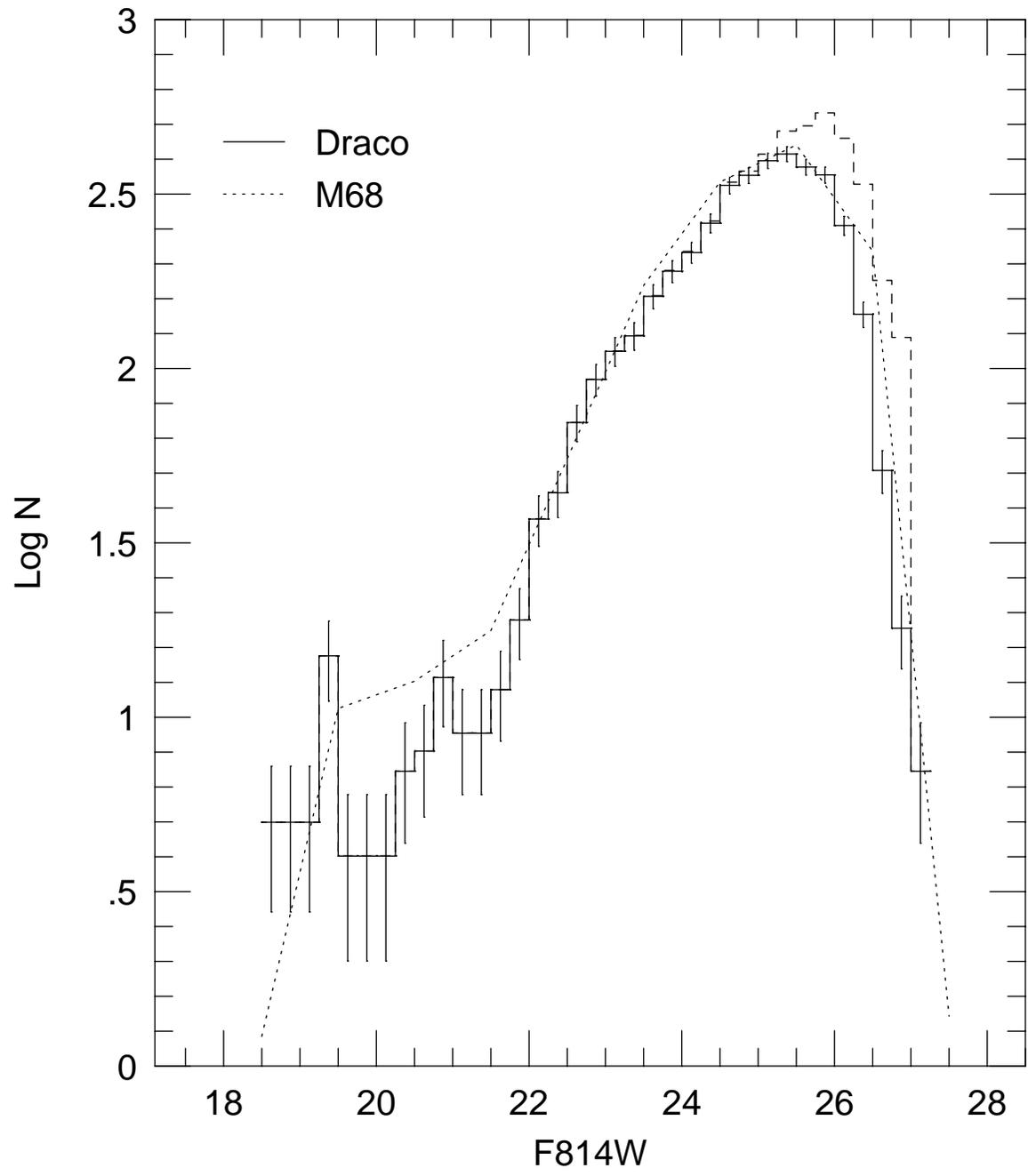
\end{document}